\documentclass[twocolumn,showpacs,preprintnumbers,amsmath,amssymb,prb]{revtex4}

\usepackage{graphicx}
\usepackage{amssymb,amsmath}
\usepackage{dcolumn}
\usepackage{bm}
\usepackage{color}
\usepackage{enumerate}
\usepackage{epstopdf}
\usepackage{ulem}

\begin{document}

\title{Datta-and-Das spin transistor controlled by a high-frequency electromagnetic field}

\author{A. S. Sheremet$^{1,2}$}
\author{O. V. Kibis$^{3,4,5}$}\email{Oleg.Kibis(c)nstu.ru}
\author{A. V. Kavokin$^{2,6,7,8}$}
\author{I. A. Shelykh$^{1,4,5}$}
\affiliation{$^1$National Research University for Information
Technology, Mechanics and Optics (ITMO), 197101 St.-Petersburg,
Russia} \affiliation{$^2$Russian Quantum Center, Novaya 100,
143025 Skolkovo, Moscow Region, Russia}
\affiliation{$^3$Department of Applied and Theoretical Physics,
Novosibirsk State Technical University, 630073 Novosibirsk,
Russia} \affiliation{$^4$Science Institute, University of Iceland
IS-107, Reykjavik, Iceland} \affiliation{$^5$Division of Physics
and Applied Physics, Nanyang Technological University, 637371
Singapore} \affiliation{$^6$School of Physics and Astronomy,
University of Southampton, SO17 1BJ, Southampton, United Kingdom}
\affiliation{$^7$Spin Optics Laboratory, St.-Petersburg State
University, 198504 Peterhof, St.-Petersburg, Russia}
\affiliation{$^8$CNR-SPIN, Tor Vergata, viale del Politechnico 1,
I-00133 Rome, Italy}


\begin{abstract}
We developed the theory of spin dependent transport through a
spin-modulator device (so-called Datta-and-Das spin transistor) in
the presence of a high-frequency electromagnetic field (dressing
field). Solving the Schr\"{o}dinger problem for dressed electrons,
we demonstrated that the field drastically modifies the spin
transport. In particular, the dressing field leads to
renormalization of spin-orbit coupling constants that varies
conductivity of the spin transistor. The present effect paves the
way for controlling the spin-polarized electron transport with
light in prospective spin-optronic devices.
\end{abstract}

\maketitle

\section{Introduction} One of most excited fields of the modern
condensed-matter physics is the physics of spin-based electronic
devices (spintronics) which is expected to play a crucial role in
the realization of high-performance information processing
\cite{Loss,Zutic,Awschalom,Cahay}. There are various kinds of
logic devices based on spin-polarized electron transport in
ferromagnets and semiconductors \cite{Siguhara,Nasirpouri,Chuang}.
In ferromagnetic materials, the spin transfer can be manipulated
by an external magnetic field. As to semiconductor structures,
their spin properties are effectively controlled with the
spin-orbit interaction. Namely, the Rashba mechanism of spin-orbit
coupling (based on the structure inversion asymmetry) and
the Dresselhaus mechanism of spin-orbit coupling (originating from
the bulk inversion asymmetry) allow tuning the spin states of
electrons in semiconductors without external magnetic fields
\cite{Koga,Ganichev,Eldridge}.

The concept of semiconductor spin-transfer device (spin
transistor) was put forward by Datta and Das in their pioneering
work \cite{Datta}. Constructively, Datta-and-Das spin transistor
consists of two magnetized ferromagnetic electrodes and a
semiconductor channel between them (see Fig.~1). In this design,
the ferromagnetic electrodes $1$ and $3$ are used to inject and
collect spin-polarized electrons, whereas the semiconductor area
$2$ serves to rotate electron spin via the spin-orbit coupling. As
a result, the transmissivity of the spin transistor depends on the
strength of the spin-orbit coupling in the semiconductor channel
\cite{Datta,Schliemann,Koo,Manchon}. Originally, the spin-orbit
coupling was proposed to be controlled with a gate voltage applied
to the semiconductor channel \cite{Datta}. In the present study,
we propose the alternative method of optical control.

It is well-known that light is an effective tool to manipulate
electronic properties of various quantum systems in the regime of
strong light-matter coupling. Since the strongly coupled system
``electron + electromagnetic field'' should be considered as a
whole, the bound electron-field object
--- ``electron dressed by electromagnetic field'' (dressed
electron)
--- became a commonly used model in modern physics
\cite{Cohen-Tannoudji_b98,Scully_b01}. It has been demonstrated
that a dressing field crucially changes physical properties of
conduction electrons in various condensed-matter structures,
including bulk semiconductors \cite{Elesin_69,Vu_04,Vu_05},
quantum wells
\cite{Mysyrovich_86,Wagner_10,Teich_13,Kibis_14,Morina_15,Pervishko_15},
quantum rings \cite{Kibis_11,Kibis_13,Joibari_14}, graphene
\cite{Lopez_08,Oka_09,Kibis_10,Kitagawa_11,Kibis_11_1,Usaj_14,Glazov_14,Lopez_15,Kristinsson_16,Kibis_16},
etc. Therefore, one can expect that spintronic devices are
strongly affected by a dressing field as well. However, a
consistent theory describing spin transport in electronic devices
subjected to electromagnetic radiation was not elaborated up to
now.

\section{Model} Let us consider a semiconductor channel of the
spin transistor irradiated by a linearly polarized electromagnetic
wave (dressing field) propagating perpendicularly to the channel
(see Fig.~1). In what follows, temperature, $T$, is assumed to be
close to zero. As a consequence, only electrons situated in the
close vicinity of the Fermi energy, $\varepsilon_F$, contribute to
conductivity of the channel.
\begin{figure}[t]
\includegraphics[width=0.95\columnwidth]{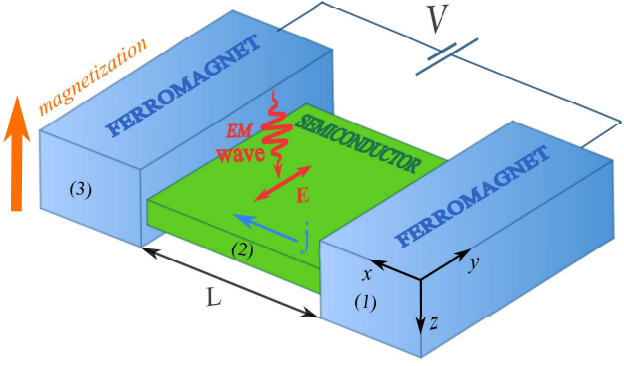}
\nopagebreak \caption{(Color online) Sketch of the spin transistor
irradiated with an electromagnetic (EM) wave. Under the applied
voltage, $V$, the source ferromagnetic contact $1$ injects the
spin-polarized current, $j$, into the semiconductor channel $2$
along the $x$ axis. The drain ferromagnetic contact $3$ collects
the current. The linearly polarized EM wave propagates
perpendicularly to the channel and its electric field, $E$, is
directed along the $y$ axis.}
 \label{sketch}
\end{figure}
The Hamiltonian of the electrons reads
\begin{eqnarray}
\label{H} \hat{\cal H}_e &=& \frac{1}{2m_s}\left(\hbar{\mathbf{k}}
- e\mathbf{A}\right)^2+\alpha\left[\bm{\sigma}\times
\left(\hbar{\mathbf{k}}-e\mathbf{A}\right)\right]_z\nonumber\\
&+&
\beta\left[\sigma_x\cdot\left(\hbar{\mathbf{k}}-{e}\mathbf{A}\right)_x
- \sigma_y\cdot\left(\hbar{\mathbf{k}}-{e}\mathbf{A}\right)_y
\right],
\end{eqnarray}
where $m_s$ is the effective electron mass in the semiconductor,
$\bm\sigma=(\sigma_x,\sigma_y,\sigma_z)$ is the Pauli matrix
vector, $\mathbf{k}=(k_x,k_y)$ is the electron wave vector,
$\mathbf{A}=(0,[E/\omega]\cos\omega t)$ is the vector potential of
the wave, $\omega$ is the wave frequency, $E$ is the amplitude of
electric field of the wave, $\alpha$ and $\beta$ are the Rashba
and Dresselhaus spin-orbit coupling constants, respectively. In
order to describe the spin transistor, we specifically consider
electrons propagating in the one-dimensional channel along the $x$
axis. Therefore, the Hamiltonian (\ref{H}) can be rewritten as
$\hat{\cal H}_e=\hat{\cal H}_0+\hat{\cal H}_k$, where
\begin{equation}
\label{H0} \hat{\cal H}_0 =
\frac{e^2E^2}{2m_s\omega^2}\cos^2\omega t + (\beta\sigma_y -
\alpha\sigma_x)\frac{eE}{\omega}\cos\omega t
\end{equation}
is the Hamiltonian describing the dressed electron state at
$\mathbf{k}=0$ and
\begin{equation}
\label{Hk} \hat{\cal H}_k = \frac{\hbar^2k_x^2}{2m_s} +
(\beta\sigma_x - \alpha\sigma_y)\hbar k_x
\end{equation}
is the Hamiltonian of ``bare'' electron in the channel. Solving
the nonstationary Schr\"odinger equation,
$i\hbar\partial\Psi_0/\partial t=\hat{\cal H}_0\Psi_0$, one can
easily obtain the exact eigenspinors of the Hamiltonian
(\ref{H0}),
\begin{equation}
\Psi_0^{\pm} = \frac{1}{\sqrt{2}}
\begin{pmatrix}
1\\ \pm \frac{\alpha - i\beta}{\gamma}
\end{pmatrix}e^{-i\frac{e^2E^2}{4m_s\hbar\omega^2}\left(t+\frac{\sin2\omega
t}{2\omega}\right)}e^{\pm i\frac{eE\gamma}{\hbar
\omega^2}\sin\omega t},
 \label{sp0}
\end{equation}
where $\gamma=\sqrt{\alpha^2 + \beta^2}$ is the effective
spin-orbit coupling constant which takes into account both the
Rashba spin-orbit interaction mechanism and the Dresselhaus one.
Since the eigenspinors (\ref{sp0}) form the complete basis of the
considered electron system, we can seek eigenspinors of the full
Hamiltonian $\hat{\cal H}_e=\hat{\cal H}_0+\hat{\cal H}_k$ as
\begin{equation}
\Psi_{k}(t) = a^{+}(t)\Psi_0^{+} + a^{-}(t)\Psi_0^{-}.
\label{solutionk}
\end{equation}
Substituting the expansion (\ref{solutionk}) into the
Schr\"{o}dinger equation with the full Hamiltonian $\hat{\cal
H}_e=\hat{\cal H}_0 + \hat{\cal H}_k$, we arrive at the two
differential equations describing the quantum dynamics of the
considered system:
\begin{eqnarray}
i\hbar\dot{a}^{\pm}(t) &=& \left[\frac{\hbar^2 k_x^2}{2m_s} \pm
\frac{2\alpha\beta}{\gamma}\hbar k_x\right] a^{\pm}(t)
\nonumber\\
&\mp & i \frac{\alpha^2-\beta^2}{\gamma}\hbar k_x e^{\mp i
\frac{2eE\gamma}{\hbar \omega^2}\sin\omega t}a^{\mp}(t).
\label{coef}
\end{eqnarray}
It follows from the Floquet theory of periodically driven quantum
systems~\cite{Zeldovich_67,Grifoni_98,Platero_04} that the wave
function (\ref{solutionk}) can be written in the form $\Psi_k(t) =
e^{-i\tilde{\varepsilon}t/\hbar}\Phi(t)$, where the function
$\Phi(t) = \Phi(t + 2\pi/\omega)$ periodically depends on time and
$\tilde{\varepsilon}$ is the quasi-energy (energy of dressed
electron). Since the quasi-energy plays the same role in
periodically driven quantum systems as the usual energy in
stationary ones, the present analysis is aimed at finding the
energy spectrum $\tilde{\varepsilon}(k_x)$ for electrons
propagating through the irradiated semiconductor channel. The
periodicity of the function $\Phi(t)$ allows seeking the
coefficients $a^{\pm}(t)$ in Eqs.~(\ref{coef}) as the Fourier
expansion,
\begin{equation}
a^{\pm}(t) = e^{-i\tilde{\varepsilon} t/\hbar}\sum_{n =
-\infty}^{\infty} a_n^{\pm} e^{i n\omega t}.
\end{equation}
As to the exponents in the right sides of Eqs.~(\ref{coef}), they
can be transformed with use of the Jacobi-Anger expansion,
$e^{iz\sin \phi} = \sum_{-\infty}^{\infty}J_n(z)e^{in\phi}$, where
$J_n(z)$ is the Bessel function of the first kind. If the dressing
field is both high-frequency and nonresonant, the rapidly
oscillating terms, $e^{i n\omega t}$, make a negligibly small
contribution into the quantum dynamics equations (\ref{coef}).
Physically, this is a general rule for periodically driven quantum
systems (see, e.g., the qualitatively similar analysis for various
dressed nanostructures in
Refs.~\cite{Kibis_14,Morina_15,Pervishko_15,Oka_09,Usaj_14}).
Therefore, the high-frequency harmonics $e^{i n\omega t}$ with
$n\neq0$ can be omitted in Eqs.~(\ref{coef}). As a consequence,
Eqs.~(\ref{coef}) can be reduced to the expression
\begin{align}
&\left(\frac{\hbar^2 k_x^2}{2m_s} \pm
\frac{2\alpha\beta}{\gamma}\hbar k_x -\tilde{\varepsilon}\right)
a_{0}^{\pm}\mp i\hbar k_x\frac{\alpha^2 - \beta^2}{\gamma}
\nonumber\\
&\times\,J_0\left(\frac{2eE\gamma}{\hbar
\omega^2}\right)a_0^{\mp}=0, \label{coef_mod}
\end{align}
Solving Eqs.~(\ref{coef_mod}), we obtain the energy spectrum of
dressed electrons,
\begin{equation}
\tilde{\varepsilon}(k_x) =  \frac{\hbar^2 k_x^2}{2m_s} \pm
\tilde{\gamma}\hbar k_x , \label{energy}
\end{equation}
where $\tilde{\gamma}=\sqrt{\tilde{\alpha}^2+\tilde{\beta}^2}$,
$\tilde{\alpha}$ and $\tilde{\beta}$ are the spin-orbit coupling
constants renormalized by a dressing field:
\begin{eqnarray}
\label{gamma} \tilde{\alpha}=\alpha
\left[J_0\left(\frac{2eE\gamma}{\hbar\omega^2}\right) +
\frac{2\beta^2}{\gamma^2}\left(1 -
J_0\left(\frac{2eE\gamma}{\hbar\omega^2}\right)\right) \right],
\nonumber\\
\tilde{\beta}=\beta
\left[J_0\left(\frac{2eE\gamma}{\hbar\omega^2}\right) +
\frac{2\alpha^2}{\gamma^2}\left(1 -
J_0\left(\frac{2eE\gamma}{\hbar\omega^2}\right)\right) \right].
\end{eqnarray}
Physically, the quantum dynamics equations (\ref{coef_mod})
exactly correspond to the stationary Schr\"odinger equation with
the effective Hamiltonian of dressed electrons in the irradiated
semiconductor channel,
\begin{equation}
\label{Hef} \hat{\cal H}_s=\frac{\hbar^2k_x^2}{2m} +
(\tilde{\beta}\sigma_x-\tilde{\alpha}\sigma_y)\hbar k_x,
\end{equation}
which results in the same energy spectrum (\ref{energy}). It
should be noted that the Hamiltonian of dressed electrons
(\ref{Hef}) exactly coincides with the well-known Hamiltonian for
"bare'' electrons (\ref{Hk}) with the formal replacement
$\alpha,\beta\rightarrow\tilde{\alpha},\tilde{\beta}$.
Consequently, all physical quantities describing dressed electrons
can be easily derived from the ``bare'' ones with the same
replacement. The energy spectrum of dressed electrons
(\ref{energy}) turns into the well-known spin-split energy
spectrum of ``bare'' electrons in the absence of the dressing
field ($E=0$). It should be stressed that the spin-orbit constants
of dressed electrons differ from the same constants for ``bare''
electrons by the Bessel-function factor depending on the dressing
field amplitude $E$ and the frequency $\omega$ [see
Eqs.~(\ref{gamma})]. In turn, the influence of the dressing field
on the renormalized constants $\tilde{\alpha}$ and $\tilde{\beta}$
depends on the ``bare'' constants ${\alpha}$ and ${\beta}$.
Particularly, the influence vanishes in the case of $|\alpha| =
|\beta|$.

\section{Spin transport} In the original proposal for the
Datta-and-Das spin transistor \cite{Datta}, the semiclassical
model was used. Within this approach, the spin-orbit interaction
was treated as an effective magnetic field which depends on the
electron wave vector along the semiconductor channel, $k_x$, and
is perpendicular to the magnetization of ferromagnetic contacts.
The spin-polarized electrons injected to the semiconductor channel
from the source ferromagnetic contact are affected by the
effective field. As a consequence, the spin of the electrons
undergoes the rotation, whose angle at the drain ferromagnetic
contact is $\theta=2m_s\gamma L/\hbar$, where $L$ is the length of
the channel. Assuming the semiconductor channel to be in the
ballistic regime, the amplitude of their probability to enter into
the drain is equal to $\cos\theta/2$. Controlling the spin-orbit
constant $\gamma=\sqrt{\alpha^2+\beta^2}$ with external actions,
one can control the conductance of the spin transistor.
Particularly, it follows from Eq.~(\ref{gamma}) that the
spin-orbit coupling can be controlled with an irradiation.
Performing the formal replacement,
$\gamma\rightarrow\tilde{\gamma}$, in the known expressions
describing the spin transistor \cite{Datta}, we obtain the
semiclassical conductance of the irradiated transistor,
\nopagebreak
\begin{equation}
G  = \left(\frac{e^2}{2h}\right)\left[1 +
\cos\left(\frac{2m_s\tilde{\gamma}L}{\hbar}\right)\right].
\label{DattaDasEquation}
\end{equation}
\nopagebreak To carry out the full quantum treatment of the spin
transport, let us divide the model structure into the regions $1$,
$2$ and $3$, where the regions $1$ and $3$ correspond to the
ferromagnetic contacts and the region $2$ corresponds to the
semiconductor channel (see Fig.~1). The electronic properties of
the irradiated semiconductor region $2$ are described by the
Hamiltonian (\ref{Hef}), whereas the Hamiltonian of ferromagnetic
regions $1$ and $3$ can be written as $\hat{\cal H}_f =
\hat{p}_x^2/2m_f-\sigma_z\Delta/2$, where $\hat{p}_x$ is the
operator of electron momentum along the $x$ axis, $m_f$ is the
electron effective mass in the ferromagnetic, and $\Delta$ is the
Zeeman spin splitting of electron states in the ferromagnetic.
Then the spinors describing the spin transistor in the three
regions, $\psi_{1,2,3}(x)$, are given by
\begin{eqnarray}
\psi_{1}(x)\left.\right|_{x<0} &=& \left(e^{iq_+x} + C_1e^{-iq_+x}\right) \begin{pmatrix}  1\\
0
\end{pmatrix} + C_2e^{q_- x} \begin{pmatrix} 0\\1 \end{pmatrix},
\nonumber\\
\psi_{2}(x)\left.\right|_{0<x<L} &=& \left(C_3e^{ik^{(1)}_+x} + C_{4}e^{ik^{(1)}_-x}\right) \begin{pmatrix} 1\\
\frac{\tilde{\beta}-i\tilde{\alpha}}{\tilde{\gamma}}
\end{pmatrix},
\nonumber\\
&+& \left(C_{5}e^{ik^{(2)}_+} x + C_6 e^{ik^{(2)}_-x}\right) \begin{pmatrix} 1\\
\frac{\tilde{\beta}+i\tilde{\alpha}}{\tilde{\gamma}} \end{pmatrix}
\nonumber\\
\psi_{3}(x)\left.\right|_{x>L} &=& C_T e^{iq_+x} \begin{pmatrix}
1\\0
\end{pmatrix} + C_7e^{- q_- x} \begin{pmatrix} 0\\1
\end{pmatrix},
\label{wave_functions}
\end{eqnarray}
where $\hbar k_{\pm}^{(1)} = \pm \sqrt{2m_s\varepsilon +
m_s^2\tilde{\gamma}^2} - m_s\tilde{\gamma}$ and $\hbar
k_{\pm}^{(2)} = \pm \sqrt{2m_s\varepsilon + m_s^2\tilde{\gamma}^2}
+ m_s\tilde{\gamma}$ are the electron momenta in the semiconductor
channel at the energy $\varepsilon$, $\hbar q_\pm =
\sqrt{{2m_f}(\Delta/2\pm\varepsilon)}$ are the electron momenta in
the ferromagnetic contacts at the same energy, and electrons in
the contacts are assumed to be fully spin polarized
($\varepsilon<\Delta/2$). In order to find the transmissivity of
the spin transistor, $C_T$, we have to use the conventional
continuity conditions at the borders of the regions \cite{Larsen},
\begin{eqnarray}
\psi_1(0)=\psi_2(0),\,\,\, \psi_2(L) = \psi_3(L),
\nonumber\\
\hat{v}_1\psi_1(0) = \hat{v}_2\psi_2(0),\,\,\, \hat{v}_2\psi_2(L)
= \hat{v}_3\psi_3(L), \label{boundary}
\end{eqnarray}
where $\hat{v}=i[\hat{\cal H},x]/\hbar$ is the velocity operator
in the different regions: $\hat{v}_{1,3} = \hat{p}_x/{m_f}$ and
$\hat{v}_2 = \hat{p}_x/{m_s}+
(\tilde{\beta}\sigma_x-\tilde{\alpha}\sigma_y)$. Substituting the
transmission amplitude, $C_T$, found from
Eqs.~(\ref{wave_functions})--(\ref{boundary}) into the Landauer
formula,
\begin{equation}
G = \frac{e^2}{h} |C_T|^2, \label{Landauer}
\end{equation}
we can calculate the quantum conductance of the spin transistor,
$G$.

\section{Discussion and conclusions} Discussing the limits of
applicability of the developed theory, we have to underline that
the considered dressing field must not be absorbable by electrons.
This is important since the absorption of radiation could be
detrimental when it generates intraband excitations (heating the
Fermi sea). It is well-known that the intraband absorption of
radiation by free conduction electrons is forbidden by the energy
conservation low and the momentum conservation low. Therefore, the
intraband absorption of THz radiation can only take place due to
the electron scattering (collisional absorption). However, the
semiconductor channel of the Datta-Das spin transistor is assumed
to be in the ballistic regime. Since scatterers are absent in the
ballistic channel, the radiation-induced intraband excitations can
be neglected. As to the optical absorption of the field by
electrons, it can be neglected if the field frequency, $\omega$,
lies far from resonant electron frequencies corresponding to
interband (intersubband) electron transitions. Therefore, the
frequency $\omega$ should be low compared to the characteristic
optical frequencies. On the other hand, the frequency, $\omega$,
should be high enough to satisfy the condition $\omega\tau\gg1$,
where $\tau\sim L/\sqrt{2\varepsilon/m_s}$ is the characteristic
transit time of electron in the channel. Physically, this
condition means that the dressing field experiences multiple
oscillations while an electron flies ballistically from the source
contact $1$ to the drain contact $3$ (see Fig.~1). Moreover, the
photon energy, $\hbar\omega$, must be larger than the
characteristic spin-orbit coupling energy,
$\varepsilon_{so}\sim{\gamma}\sqrt{2m_s\varepsilon}$, in which
case it is safe to neglect the rapidly oscillating terms in the
quantum dynamics equations (\ref{coef}). Therefore, the dressing
field must be both high-frequency and non-resonant.
\begin{figure}[t]
\includegraphics[width=0.95\columnwidth]{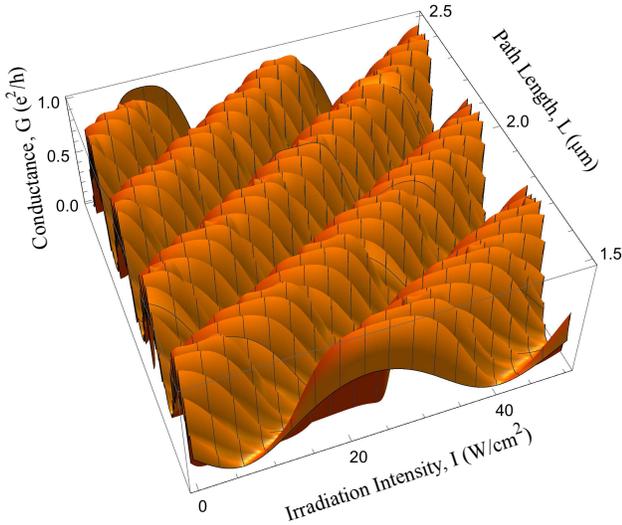}
\caption{(Color online) The quantum conductance of the irradiated
spin transistor, $G$, as a function of the irradiation intensity,
$I$, and the length of the semiconductor channel (path length),
$L$, for the Rashba and Dresselhaus spin-orbit coupling constants
$\alpha =\beta/4 = 5\cdot 10^3$ m/s.}
 \label{Plot3D}
\end{figure}
\begin{figure}[t]
\includegraphics[width=0.95\columnwidth]{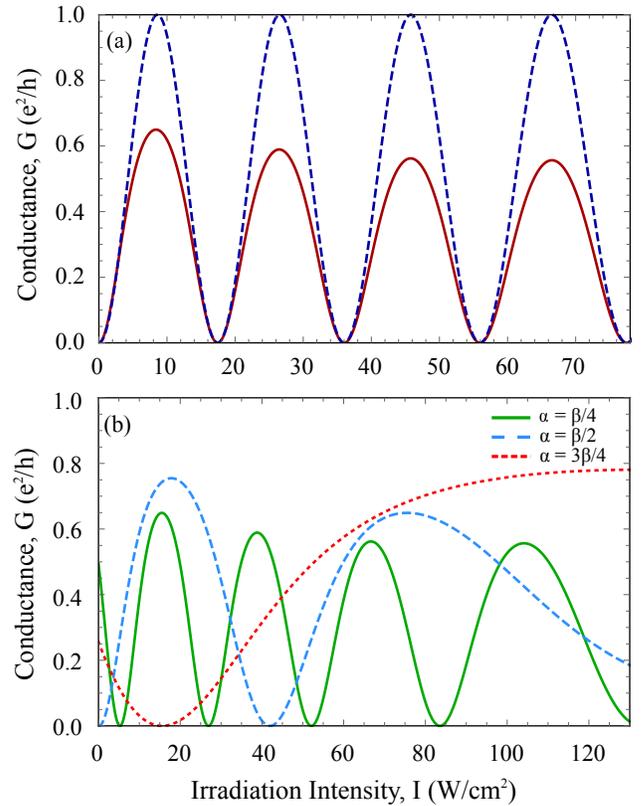}
\caption{(Color online) The conductance of spin transistor, $G$,
as a function of the irradiation intensity, $I$, for the length of
the semiconductor channel $L = 2.3\,\mu$m: (a) semiclassical
conductance (dashed line) and quantum conductance (solid line) for
the Dresselhaus constant $\beta = 2\cdot 10^4$ m/s and the Rashba
constant $\alpha=0$; (b) quantum conductance for the Dresselhaus
constant $\beta = 2\cdot 10^4$ m/s and the different Rashba
constants, $\alpha$.}
 \label{conductance}
\end{figure}

At low temperatures, the electrons contributing to the spin
transport in the channel are situated at the Fermi level.
Therefore, the energy $\varepsilon$ in all expressions above
coincides with the Fermi energy in the semiconductor channel,
$\varepsilon_F$. The temperature, $T$, should be low compared to
both the Fermi energy, $\varepsilon_F$, and the spin-orbit
coupling energy,
$\varepsilon_{so}\sim{\gamma}\sqrt{2m_s\varepsilon_F}$. It should
be noted also that employing the concept of dressing field we
assume that $\hbar/\tau_{\varphi}<{g}_0$ , where $\tau_{\varphi}$
is the characteristic dephasing time, and
${g}_0=eE{\gamma}/\omega$ is the characteristic constant of
light-matter coupling in the semiconductor channel. In the
considered case of ballistic channel, the dephasing time can be
estimated as $\tau_\varphi\sim\tau\sim L/\sqrt{2\varepsilon/m_s}$.
Therefore, the field amplitude, $E$, should be large enough to
satisfy the condition
$E>\hbar\omega\sqrt{2\varepsilon/m_s}/(e\gamma L)$.

Numerical calculations of the conductance of the irradiated spin
transistor, $G$, with use of the semiclassical Datta-and-Das
formula (\ref{DattaDasEquation}) and the quantum Landauer formula
(\ref{Landauer}) are presented in Figs.~2 and 3 for the field
frequency $\omega = 1$ THz, the Fermi energy $\varepsilon_F = 10$
meV, the Zeeman splitting in ferromagnetic contacts $\Delta = 40$
meV, the electron effective masses $m_f = m_e$ (in the
ferromagnetic contacts) and $m_s = 0.067m_e$ (in the GaAs
semiconductor channel), where $m_e$ is the mass of free electron.
The difference between the quantum and semiclassical results is
due to the multiple electron scattering between the source and
drain contacts, which is not accounted for within the
semiclassical approach. The multiple scattering leads to formation
of the standing electron waves between the contacts, which have
the same physical nature as eigenmodes of the Fabry-Perot
resonator. As a consequence, the amplitude of quantum conductance
experiences additional modulation (see Fig.~3).

The plotted dependence of the conductance, $G$, on the irradiation
intensity, $I$, originates from the field-induced modification of
the spin-orbit coupling constants (\ref{gamma}). Conventionally,
only the gate voltage was used before to tune the constants
\cite{Manchon,Nitta,Heida,Sadreev_13}. In particular, the
gate-controlled scheme of the spin transistor was proposed by
Datta and Das \cite{Datta}. The present theory indicates that the
spin-orbit coupling can be effectively controlled with an external
optical field as well.

To summarize, the strong coupling between electron spins and a
high-frequency electromagnetic field is an efficient control tool
for the spin precession rate in spin transistors. Since the
dressing field renormalizes spin-orbit coupling constants, the
variation of the dressing field intensity drastically modifies the
spin transport properties of the transistors. Thus, the predicted
effect opens the way to realization of new spin-optronic devices.
Since light-controlled electronic devices are typically much
faster than those of electrically controlled, the
optically-controlled Datta-and-Das spin transistor is expected to
be faster than the gate-controlled one.

\begin{acknowledgments} This work was partially supported by FP7
IRSES projects POLATER and QOCaN, the Rannis project BOFEHYSS,
Singapore Ministry of Education under AcRF Tier 2 grant
MOE2015-T2-1-055, the Russian Target Federal Program ``Research
and Development in Priority Areas of Development of the Russian
Scientific and Technological Complex for 2014--2020'' (project
RFMEFI58715X0020). ASS thanks the University of Iceland for
hospitality.
\end{acknowledgments}


\begin{thebibliography}{99}

\bibitem{Loss} D. D. Awschalom, D. Loss, N. Samarth,
\textit{Semiconductor Spintronics and Quantum Computation}
(Springer-Verlag, Berlin, 2002).

\bibitem{Zutic} I. Zutic, J. Fabian and S. D. Sarma,
Rev. Mod. Phys. \textbf{76}, 323 (2004).

\bibitem{Awschalom} D. D. Awschalom and M. E. Flatt\'{e},
Nat. Phys. \textbf{3}, 153 (2007).

\bibitem{Cahay} M. Cahay, Nat. Nanotechnol. \textbf{10}, 21 (2015).

\bibitem{Siguhara} S. Sugahara, J. Nitta, Proc. IEEE \textbf{98}, 2124 (2010).

\bibitem{Nasirpouri} F. Nasirpouri and A. Nogaret,
\textit{Nanomagnetism and Spintronics: Fabrication, Materials,
Characterization and Application} (World Scientific, Singapore,
2011).

\bibitem{Chuang} P. Chuang, S.-C. Ho, L. W. Smith, F. Sfigakis, M. Pepper,
C.-H. Chen, J.-C. Fan, J. P. Griffiths, I. Farrer, H. E. Beere, G.
A. C. Jones, D. A. Ritchie, T.-M. Chen, Nat. Nanotechnol.
\textbf{10}, 35 (2015).

\bibitem{Koga} T. Koga, J. Nitta, T. Akazaki, H. Takayanagi,
Phys. Rev. Lett. \textbf{89}, 046801 (2002).

\bibitem{Ganichev} S. D. Ganichev, V. V. Bel'kov,
L. E. Golub, E. L. Ivchenko, P. Schneider, S. Giglberger, J.
Eroms, J. De Boeck, G. Borghs, W. Wegscheider, D. Weiss, W.
Prettl, Phys. Rev. Lett. \textbf{92}, 256601 (2004).

\bibitem{Eldridge} P. S. Eldridge, W. J. H. Leyland,
P. G. Lagoudakis, O. Z. Karimov, M. Henini, D. Taylor, R. T.
Phillips, R. T. Harley, Phys. Rev. B \textbf{77}, 125344 (2008).

\bibitem{Datta} S. Datta, B. Das,
Appl. Phys. Lett. \textbf{56}, 665 (1990).

\bibitem{Schliemann} J. Schliemann, J. C. Egues, D. Loss,
Phys. Rev. Lett. \textbf{90}, 146801 (2003).

\bibitem{Koo} H. C. Koo, J. H. Kwon, J. Eom, J. Chang,
S. H. Han, M. Johnson, Science \textbf{325}, 1515 (2009).

\bibitem{Manchon} A. Manchon, H.C. Koo, J. Nitta, S. M. Frolov,
R. A. Duine, Nat. Mat. \textbf{14}, 871 (2016).

\bibitem{Cohen-Tannoudji_b98}
C. Cohen-Tannoudji, J. Dupont-Roc, G. Grynberg,
\textit{Atom-Photon Interactions: Basic Processes and
Applications}  (Wiley, Chichester, 1998).

\bibitem{Scully_b01}
M. O. Scully, M. S. Zubairy, \textit{Quantum Optics} (Cambridge
University Press, Cambridge, 2001).

\bibitem{Elesin_69}
S. P. Goreslavskii, V. F. Elesin, JETP Lett. {\bf 10}, 316 (1969).

\bibitem{Vu_04}
Q. T. Vu, H. Haug, O. D. M\"ucke, T. Tritschler, M. Wegener, G.
Khitrova, H. M. Gibbs, Phys. Rev. Lett. \textbf{92}, 217403
(2004).

\bibitem{Vu_05}
Q. T. Vu, H. Haug, Phys. Rev. B {\bf 71}, 035305 (2005).

\bibitem{Mysyrovich_86}
A. Myzyrowicz, D. Hulin, A. Antonetti, A. Migus, W. T. Masselink,
H. Morko\c{c}, Phys. Rev. Lett. {\bf 56}, 2748 (1986).

\bibitem{Wagner_10}
M. Wagner, H. Schneider, D. Stehr, S. Winnerl, A. M. Andrews, S.
Schartner, G. Strasser, M. Helm, Phys. Rev. Lett. {\bf 105},
167401 (2010).

\bibitem{Teich_13}
M. Teich, M. Wagner, H. Schneider, M. Helm, New J. Phys. {\bf 15},
065007 (2013).

\bibitem{Kibis_14}
O. V. Kibis, Europhys. Lett. {\bf 107}, 57003 (2014).

\bibitem{Morina_15}
S. Morina, O. V. Kibis, A. A. Pervishko, I. A. Shelykh, Phys. Rev.
B {\bf 91}, 155312 (2015).

\bibitem{Pervishko_15}
A. A. Pervishko, O. V. Kibis, S. Morina, I. A. Shelykh, Phys. Rev.
B {\bf 92}, 205403 (2015).

\bibitem{Kibis_11}
O. V. Kibis, Phys. Rev. Lett. {\bf 107}, 106802 (2011).

\bibitem{Kibis_13} O. V. Kibis, O. Kyriienko, I. A. Shelykh, Phys. Rev. B {\bf 87},
245437 (2013).

\bibitem{Joibari_14}
F. K. Joibari, Y. M. Blanter, G. E. W. Bauer, Phys. Rev. B {\bf
90}, 155301 (2014).

\bibitem{Lopez_08}
F. J. L\'{o}pez-Rodr\'{i}guez, G. G. Naumis, Phys. Rev. B {\bf
78}, 201406(R) (2008).

\bibitem{Oka_09}
T. Oka and H. Aoki, Phys. Rev. B {\bf 79}, 081406(R) (2009).

\bibitem{Kibis_10} O. V. Kibis, Phys. Rev. B {\bf 81}, 165433
(2010).

\bibitem{Kitagawa_11}
T. Kitagawa, T. Oka, A. Brataas, L. Fu, E. Demler, Phys. Rev. B
{\bf 84}, 235108 (2011).

\bibitem{Kibis_11_1}
O. V. Kibis, O. Kyriienko, I. A. Shelykh, Phys. Rev. B {\bf 84},
195413 (2011).

\bibitem{Usaj_14}
G. Usaj, P. M. Perez-Piskunow, L. E. F. Foa Torres, C. A.
Balseiro, Phys. Rev. B {\bf 90}, 115423 (2014).

\bibitem{Glazov_14}
M. M. Glazov, S. D. Ganichev, Phys. Rep. {\bf 535}, 101 (2014).

\bibitem{Lopez_15}
A. L\'{o}pez, A. Di Teodoro, J. Schliemann, B. Berche, B. Santos,
Phys. Rev. B {\bf 92}, 235411 (2015).

\bibitem{Kristinsson_16}
K. Kristinsson, O. V. Kibis, S. Morina, I. A. Shelykh, Sci. Rep.
{\bf 6}, 20082 (2016).

\bibitem{Kibis_16}
O. V. Kibis, S. Morina, K. Dini, I. A. Shelykh, Phys. Rev. B {\bf
93}, 115420 (2016).

\bibitem{Zeldovich_67}
Ya. B. Zel'dovich, Sov. Phys. JETP {\bf 24}, 1006 (1967).

\bibitem{Grifoni_98}
M. Grifoni, P. H\"anggi, Phys. Rep. {\bf 304}, 229 (1998).

\bibitem{Platero_04}
G. Platero, R. Aguado, Phys. Rep. {\bf 395}, 1 (2004).

\bibitem{Larsen} M. H. Larsen, A. M. Lunde, K. Flensberg,
Phys. Rev. B \textbf{66}, 033304 (2002).

\bibitem{Nitta} J. Nitta, T. Akazaki, H. Takayanagi, T. Enoki, Phys. Rev. Lett. \textbf{78}, 1335 (1997).

\bibitem{Heida} J. P. Heida, B. J. van Wees, J. J. Kuipers, T. M. Klapwijk, G. Borghs, Phys. Rev. B \textbf{57}, 11911 (1998).

\bibitem{Sadreev_13} A. F. Sadreev, E. Y. Sherman, Phys. Rev. B {\bf 88}, 115302
(2013).

\end{thebibliography}
\end{document}